\documentclass[english,twocolumn,aps,prl]{revtex4}
\usepackage[T1]{fontenc}
\usepackage[latin1]{inputenc}
\usepackage{amsmath}
\usepackage{graphicx}
\usepackage{amssymb}
\usepackage{bm}
\usepackage{hyperref}
\usepackage{hypernat}
\usepackage{epsfig}
\usepackage{amsfonts}
\usepackage{amsthm}
\usepackage{babel}

\begin{document}

\title{Paired states in spin-imbalanced atomic Fermi gases in one dimension}


\author{Paata Kakashvili and C. J. Bolech}

\affiliation{Department of Physics and Astronomy, Rice University, Houston, Texas
77005}

\begin{abstract}
A growing expertise to engineer, manipulate and probe different cold-atom
analogs of electronic condensed matter systems allows to probe properties
of exotic pairing. We study paired states of spin-imbalanced ultracold
atomic system of fermions with attractive short-range interactions
in one-dimensional traps. Calculations are done using the Bethe Ansatz
technique and the trap is incorporated into the solution via a local
density approximation. The thermodynamic-Bethe-Ansatz equations are
solved numerically and different local density profiles are calculated
for zero and finite temperatures. A procedure to identify the homogeneous-system
phase diagram using local density profiles in the trap is also proposed.
Such scheme would be immediately useful for the experiments. 
\end{abstract}
\maketitle

\begin{figure}[!htb]
\includegraphics[width=3.3in]{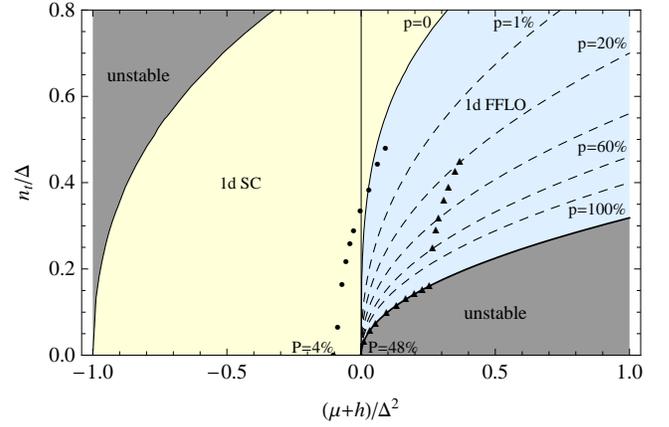} 
\caption{Zero-temperature phase digram of the uniform 1d system,
showing BCS-like (1d SC), FFLO-like (1d FFLO) and (fully polarized)
normal ($p=100\%$ curve) ``phases''. The areas marked ``unstable''
correspond to parameter regimes, which do not support equilibrium
states. Local density trap profiles for total polarizations $P=4\%$
(circles) and $P=48\%$ (triangles) are superimposed on the diagram.}
\label{PhaseDiagram} 
\end{figure}

Recently, ultracold-atom realizations of condensed-matter-physics
models have attracted a lot of interest. A growing expertise in engineering,
tuning and probing atomic systems allows to study exceedingly complicated
systems. One of the current challenges of condensed matter physics
is to understand the different exotic paired states that are realized
when paired particles have different chemical potentials. This disparity
could be due to different physical effects such as different masses
of the pairing particles or external magnetic filed and is thought
to arise in systems such as quasi one-dimensional and two-dimensional
organic superconductors~\cite{Organic_FFLO} and heavy-fermion materials~\cite{Heavy_Fermion_FFLO}
to name a few. It is argued that the zero-center-of-mass-momentum
Cooper pairs are destabilized due the population imbalance and different,
unconventional, paired states are proposed to occur~\cite{sarma_influence_1963,liu_interior_2003}.
Among these exotic states, the FFLO (Fulde-Ferrell-Larkin-Ovchinnikov)
state, in which pairs have nonzero center-of-mass momentum and superconductivity
(superfluidity) coexists with nonzero polarization, is one of the
candidates~\cite{fulde_superconductivity_1964,larkin_inhomogeneous_1965}.
Recently, several experiments have studied the interplay of pairing
and polarization in tree-dimensional (3d) spin-imbalanced superfluid
clouds~\cite{zwierlein_fermionic_2006_short_combined,partridge_pairing_2006_short_combined}.
Ongoing experiments are exploring different pairing mechanisms in
spin-imbalanced ultracold atomic systems of fermions in reduced dimensions~\cite{Hulet_2008},
where the FFLO state is believed to have a large parameter regime
of stability. The ultracold cloud of atoms is subjected to a 2d optical
lattice, which defines an array of quasi-1d tubes, which can be regarded
as independent if the intensity of the laser beams is large enough
to suppress tunneling among them. Thus, to explore pairing properties
of 1d systems of spin-imbalanced fermions is very relevant. There
have been many different recent studies of this system. Bosonization~\cite{yang_inhomogeneous_2001,zhao_theory_2008},
Bethe Ansatz~\cite{orso_attractive_2007,hu_phase_2007,guan_phase_2007},
mean field~\cite{liu_fulde-ferrell-larkin-ovchinnikov_2007_combined},
QMC~\cite{batrouni_exact_2008,casula_quantum_2008} and DMRG~\cite{feiguin_pairing_2007,rizzi_fulde-ferrell-larkin-ovchinnikov_2008_short,tezuka_density-matrix_2008,feiguin_spectral_2008}
methods have been used to study this problem and the agreement is
that the partially polarized superfluid region (or {}``phase'')
is the analog of the FFLO state in higher dimensions (see Fig.~\ref{PhaseDiagram},
explained below). Most of the past analysis was done for zero temperature,
except the QMC and mean field calculations. In addition, these approaches
find the regime of strong interactions between particles very challenging,
while experiments are done preferably in that regime. Therefore, there
is a need for finite-temperature calculations, which do not suffer
of the above limitations. 

In this paper, we address the problem using the Thermodynamic-Bethe-Ansatz
(TBA) equations~\cite{takahashi__ground_1970,takahashi_one-dimensional_1971},
which provide a full finite-$T$ description and are well posed for
arbitrary interaction strengths. We also propose a scheme to reconstruct
the ``phase diagram'' of a homogeneous 1d attractive fermion cloud
from experimental profiles of local particle density and polarization
in a trap. We start from the Hamiltonian of a single 1d tube in a
harmonic trap
\begin{equation}
H=-\frac{\hbar^{2}}{2m}\sum_{i}\frac{\partial^{2}}{\partial z_{i}^{2}}+g_{\text{1D}}\sum_{i<j}\delta(z_{i}-z_{j})+\frac{m\omega_{z}^{2}}{2}\sum_{i}z_{i}^{2},
\label{Hamiltonian1}
\end{equation}
where $m$ is the mass of the atoms, $\omega_{z}$ is the trap frequency
and $g_{\text{1D}}=\frac{2\hbar^{2}a_{\text{3D}}}{ma_{\perp}^{2}}\frac{1}{1-Ca_{\text{3D}}/a_{\perp}}$
[with $C=|\zeta(1/2)|/\sqrt{2}\simeq1.0326$] is the effective interaction
strength between particles, which can be tuned using a Feshbach resonance
and made attractive ($g_{\text{1D}}<0$)~\cite{olshanii_atomic_1998}.
$a_{\text{3D}}$ is the 3d scattering length and $a_{\perp}$ is the
transverse oscillator length. In what follows, we shall measure lengths
in units of the harmonic oscillator length $(\hbar/m\omega_{z})^{1/2}$
and energies in units of $\hbar\omega_{z}/2$. The dimensionless Hamiltonian
reads 
\begin{equation}
H=-\sum\frac{\partial^{2}}{\partial z_{i}^{2}}-4\Delta\sum\delta(z_{i}-z_{j})+\sum z_{i}^{2},
\label{Hamiltonian2}
\end{equation}
where $\Delta=-\frac{g_{\text{1D}}}{2\hbar\omega_{z}}\sqrt{\frac{m\omega_{z}}{\hbar}}>0$
(attractive interactions). We will use the solution of the uniform
system and take trap effects into account via the local density (Thomas-Fermi)
approximation. This approximation is widely used in studying trapped
gases, but its violation has been observed in highly elongated 3d
clouds~\cite{partridge_pairing_2006_short_combined} due to interface
interaction effects~\cite{de_silva_surface_2006,imambekov_breakdown_2006}.
In 1d, one rather finds smooth crossovers and, for attractive fermions,
comparison with QMC calculations~\cite{casula_quantum_2008} indicate
that the local density approximation gives good results for density
profiles. Most features of local trap profiles are faithfully reproduced,
but for the short wavelength Friedel oscillations, which are washed
out at finite temperatures. In this framework, the Bethe Ansatz solution
of the model~\cite{yang_exact_1967,gaudin_un_1967} can be used \emph{locally}.
Eigenvalues and eigenstates of the system (of size $L$ and with periodic
boundary conditions) are determined from the \emph{roots} of the Bethe-Ansatz
equations: 
\begin{eqnarray*}
e^{ik_{j}L} & = & \prod_{m=1}^{M}\frac{k_{j}-\Lambda_{m}-i\Delta}{k_{j}-\Lambda_{m}+i\Delta},\\
-\prod_{m=1}^{M}\frac{\Lambda_{n}-\Lambda_{m}-2i\Delta}{\Lambda_{n}-\Lambda_{m}+2i\Delta} & = & \prod_{j=1}^{N}\frac{\Lambda_{n}-k_{j}-i\Delta}{\Lambda_{n}-k_{j}+i\Delta},
\end{eqnarray*}
where $j=1,...,N$ and $n=1,...,M$ ($M\le N/2$), with $N$ the total
number of particles and $M$ the number of minority ones. $\{ k_{i}\}$
are particle quasi-momenta and solely determine energy eigenvalues,
$E=\sum_{j=1}^{N}k_{j}^{2}\equiv eL$. $\{\Lambda_{n}\}$ are (effective)
\emph{spin} rapidities and determine \emph{magnetic} properties of
the system. Analyzing the roots of the Bethe Ansatz equations one
can find that there are three classes of solutions~\cite{takahashi_one-dimensional_1971}:
real $k_{j}$-s, representing unpaired particles; bound states of
two particles with opposite spins, represented by complex $k_{j}=\Lambda_{j}\pm i\Delta$,
with real spin rapidity $\Lambda_{j}$; and complex spin rapidities
forming strings $\Lambda^{(r)j}=\Lambda^{(r)}+i\Delta(r+1-2j)$, with
$j=1,...,r$ and with real $\Lambda^{(r)}$ the center of the $r$-string.

In the thermodynamic limit, $L\rightarrow\infty$ (keeping $N/L$
and $M/L$ constant), it is possible to \emph{locally} define root
and hole density distributions for unpaired particles, bound states
and $n$-strings: $\rho_{u}^{r(h)}(k,z)$, $\rho_{b}^{r(h)}(k,z)$
and $\rho_{sn}^{r(h)}(\Lambda,z)$. The partition function is defined
as $Z=\text{Tr}e^{-H/T}=e^{-F/T}$, where $\frac{F}{L}={\cal F}(z)=e(z)-Ts(z)-\mu(z)n_{t}(z)-hn_{s}(z)$
is the free energy density, $\mu(z)=\mu_{0}-z^{2}$ is the \emph{local}
chemical potential, $h$ is the applied magnetic field, and $s(z)$,
$n_{t}(z)$ and $n_{s}(z)$ are local entropy, total-particle and
spin densities, respectively. Then the equilibrium root and hole density
distributions are locally determined by minimizing the free energy
while taking into account the Bethe-Ansatz equations. The resulting
TBA equations~\cite{takahashi__ground_1970,takahashi_one-dimensional_1971}
are 
\begin{eqnarray*}
 &  & f_{u}-\bar{f}_{u}=G*f_{b}-G*f_{s1},\\
 &  & f_{b}-\bar{f}_{b}=2[k^{2}-\Delta^{2}-\mu(z)]/T+K_{1}*\bar{f}_{u}+K_{2}*\bar{f}_{b},\\
 &  & f_{sn}-\bar{f}_{sn}=\delta_{n,1}G*\bar{f}_{u}+G*(f_{sn+1}+\hat{\delta}_{n,1}f_{sn-1}),\\
 &  & \lim_{n\rightarrow\infty}(K{}_{n+1}*f_{sn}-K{}_{n}*f_{sn+1})=-2h/T,
\end{eqnarray*}
where $f_{\alpha}=\ln(1+\rho_{\alpha}^{h}/\rho_{\alpha}^{r})$ and
$\bar{f}_{\alpha}=\ln(1+\rho_{\alpha}^{r}/\rho_{\alpha}^{h})$ are
functions related to probability distributions, $K_{n}(k)=\frac{1}{\pi}\frac{n\Delta}{k^{2}+(n\Delta)^{2}}$,
$G(k)=\frac{1}{4\Delta\cosh\frac{\pi k}{2\Delta}}$ and `$*$' indicates
the convolution of two functions (these notations are as in Ref.~\cite{bolech_solution_2005}).
In terms of the $f$'s, ${\cal F}(z)$ reads 
\begin{equation}
{\cal F}(z)=-\frac{T}{2\pi}\int dk\bar{f}_{u}(k,z)-\frac{T}{\pi}\int dk\bar{f}_{b}(k,z).\label{FreeEnergy}
\end{equation}

It is possible to straightforwardly determine densities of different
thermodynamic variables from the free energy using $x(z)=-\frac{\partial{\cal F}(z)}{\partial\bar{x}(z)}$,
where $x=n_{t},n_{s},s$ and $\bar{x}=\mu,h,T$ are conjugate thermodynamic
variables. One can write equations for the different density distributions
of all extensive thermodynamic variables in a compact form
\begin{eqnarray}
 &  & \rho_{u}^{x,h}+\rho_{u}^{x,r}=G*\rho_{b}^{x,h}+G*\rho_{s1}^{x,h},\label{DENS}\\
 &  & \rho_{b}^{x,h}+\rho_{b}^{x,r}=\delta_{x,n_{t}}/\pi-K_{1}*\rho_{u}^{x,r}-K_{2}*\rho_{b}^{x,r},\nonumber \\
 &  & \rho_{sm}^{x,h}+\rho_{sm}^{x,r}=\delta_{m,1}G*\rho_{u}^{x,r}+G*(\rho_{sm+1}^{x,h}+\hat{\delta}_{m,1}\rho_{sm-1}^{x,h}),\nonumber \\
 &  & \lim_{m\rightarrow\infty}(K{}_{n+1}*\rho_{sm}^{x,h}-K{}_{n}*\rho_{sm+1}^{x,h})=-\delta_{x,n_{s}}/\pi,\nonumber 
\end{eqnarray}
where the corresponding density distributions are defined as $\rho_{y}^{x,r}=(-1)^{\delta_{y,s}}\frac{1}{2\pi}\partial_{\bar{x}}(T\bar{f}_{y})$
and $\rho_{y}^{x,h}=(-1)^{1+\delta_{y,s}}\frac{1}{2\pi}\partial_{\bar{x}}(Tf_{y})$.
Notice that the above system of equations for the particle density
($x=n_{t}$) coincides with the continuum limit of the Bethe-Ansatz
equations. Using these definitions, local densities of thermodynamic
observables in the trap are given by 
\begin{equation}
x(z)=\int dk\rho_{u}^{x,r}(k,z)+2\int dk\rho_{b}^{x,r}(k,z).
\label{DENSITY}
\end{equation}
\begin{figure*}[t]
\includegraphics[width=7in]{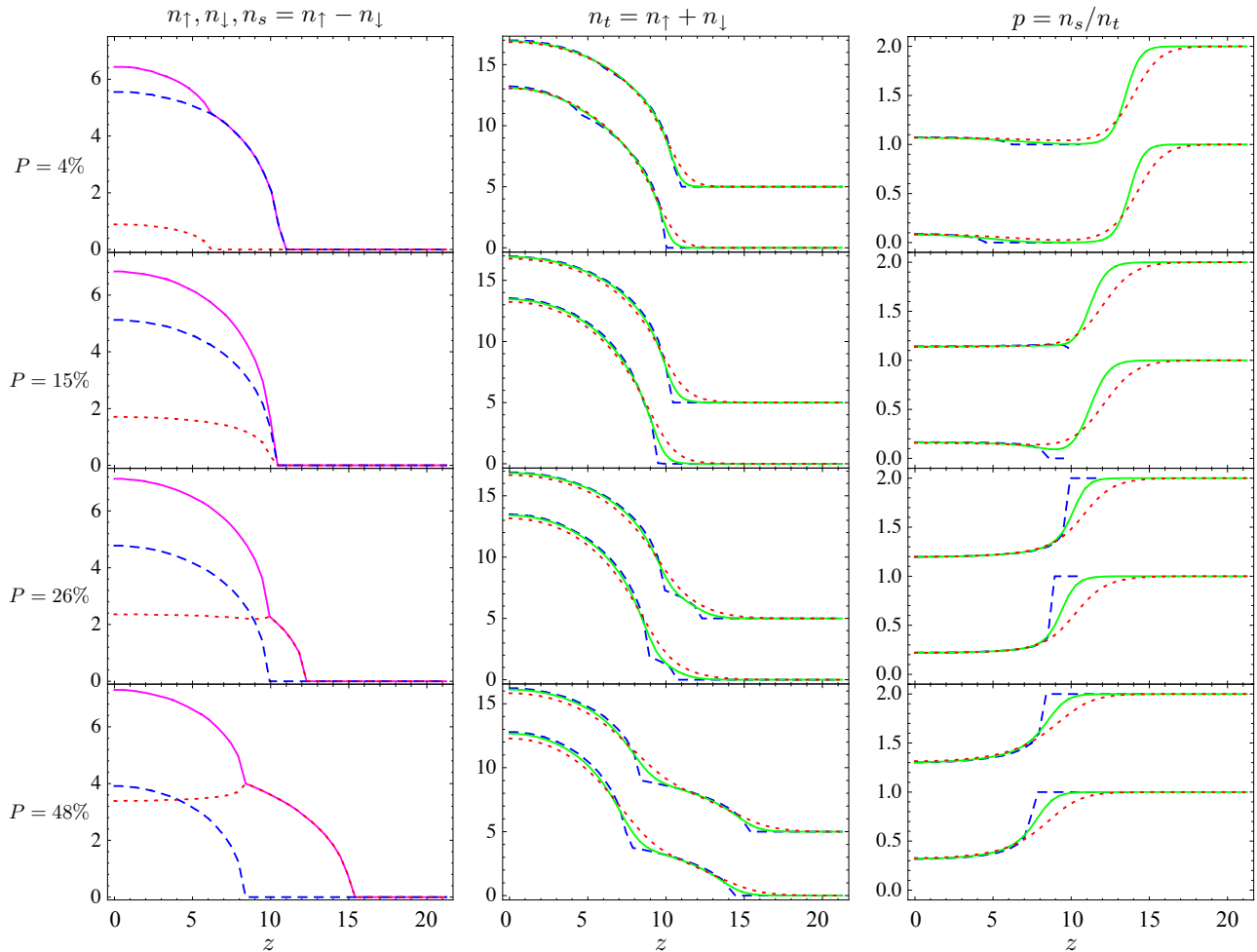} 
\caption{Trap density profiles for $N=200$, $T=0,0.06,0.12[\mu_{\uparrow F}]$
(dashed, solid and dotted, respectively) and different total polarization
$P$. The first column shows spin-up ($n_{\uparrow}$, solid), spin-down
($n_{\downarrow}$, dashed) particle densities and their difference
($n_{s}$, dotted) for $\Delta=25$ and $T=0$ only. The second column
shows the local particle densities ($n_{t}$) for $\Delta=25,100$
and different temperatures. The third column shows the respective
local polarizations ($p$). The $n_{t}$ and $p$ plots for $\Delta=25$
are vertically offset for clarity.}
\label{TrapProfiles} 
\end{figure*}

The integral equations are solved numerically for different temperatures
and interaction strengths (cf. Ref.~\cite{bolech_solution_2005});
different density profiles are plotted in Fig.~\ref{TrapProfiles}.
The reference chemical potential, $\mu_{0}$, and the magnetic field,
$h$, are fixed by the self-consistency relations $\int dz\, n_{t}(z)=N$
and $\int dz\, n_{s}(z)=PN$. Depending on $P$, trap profiles show
different regimes that reveal properties of the zero-temperature ``phase
digram'' of a uniform system. For small polarizations ($P=4\%$ plots),
an FFLO-like state in the trap center coexists with a BCS-like state
at the edges; while for large polarizations ($P=26\%$ and $P=48\%$
plots), the FFLO-like state in the trap center coexists with a fully
polarized normal state at the edges. Close to the ``critical point''
($P=15\%$ plots), the entire trap is in the FFLO-like state. The
interface between different ``phases'' in the trap is marked by
a kink feature in the local particle density (or polarization) profiles
at zero temperature, which is smeared out by thermal fluctuations.
From Fig.~\ref{TrapProfiles} we see that while for cases with large
total polarization the phase boundary is still visible at $T=0.06\,[\mu_{\uparrow F}]$
(here $\mu_{\uparrow F}$ is the noninteracting Fermi energy for the
majority species), for small polarizations temperature should be even
smaller to observe the kink. From the local polarization profiles,
we see that in the normal state temperature induces bound pairs, while
in the fully paired state unpaired particles appear.

We propose here a scheme to identify the low-temperature ``phase
diagram'' of the uniform system from trap profiles of the local particle
density and polarization. To achieve that aim we represent the phase
diagram in variables that can be measured directly in the experiments.
We argue that $n_{t}(z)$ and the majority-spin local chemical potential
$\mu_{\uparrow}(z)=\mu(z)+h$ can be straightforwardly measured and,
thus, a ``phase diagram'' in these variables should be directly
useful to analyze experimental findings. In these variables, the ``phase
diagram'' of the uniform system (see Fig.~\ref{PhaseDiagram}) shows
three different states: the BCS-like fully paired state with zero
polarization (1dSC), the FFLO-like state with non-zero polarization
(1dFFLO) and the fully polarized normal state which has collapsed
into a curve (cf.~Fig.~1 in Ref.~\cite{orso_attractive_2007}).
Plotting the local particle density and polarization profiles as functions
of $\mu_{\uparrow}$ will reconstruct the uniform-system {}``phase
diagram'' and probe \emph{into} the 1dFFLO state. To elucidate the
scheme we consider a trap profile with the total polarization $P=48\%$
(triangles on Fig.~\ref{PhaseDiagram}). In this regime, the 1dFFLO 
``phase'' in the trap center coexists with the fully polarized
region at the edges (see Fig.~\ref{TrapProfiles}). In the framework
of the local density approximation, the radius of the cloud is given
by $R=\sqrt{\mu_{0}+h}$ and thus one can easily find $\mu_{\uparrow}(z)+h=R^{2}-z^{2}$.
Thus, the trap profile can be uniquely \emph{placed} on the ``phase
digram'' (see Fig.~\ref{PhaseDiagram}). The above identification
is valid for clouds with fully polarized wings and plotting many profiles
of these type allows to probe \emph{into} the 1dFFLO. For small total
polarizations, the 1dFFLO in the trap center coexists with the 1dSC
regions at the edges (see Fig.~\ref{TrapProfiles}). In this case,
the radius of the cloud is given by $R=\sqrt{\Delta^{2}+\mu_{0}}$
and thus one can easily find $\mu_{\uparrow}(z)+h=R^{2}+h-\Delta^{2}-z^{2}$.
It is again possible to place a trap profile on the {}``phase digram''
(see trap profile for $P=4\%$ in Fig.~\ref{PhaseDiagram}). Due
to the nonzero temperature and measurement noise, it will be challenging
to precisely identify the zero polarization curve (1dFFLO-1dSC boundary),
but most features of the ``phase diagram'' (including inner 1dFFLO
isopolarization lines) are, nevertheless, within the reach of the
upcoming experiments.

To summarize, we have studied different pairing states in spin-imbalanced
ultracold atomic clouds of fermions in 1d. We have combined the Bethe
Ansatz method and the local density approximation to calculate trap
profiles of different observables. Our full finite-temperature calculations
shed light on the parameter regimes for total density, polarization
and temperature that are required to observe exotic states such as
the 1dFFLO. We also proposed a way to unambiguously reconstruct the 
``phase'' diagram of the uniform system from experimental local
density profiles of the trapped clouds. Our calculational scheme could
also be used to compute asymptotic momentum distributions that could
be measured after simultaneously releasing the trap and switching
off the interactions in time-of-flight experiments. This would be
complementary to the in-situ pair-momentum correlators discussed already
in the literature~\cite{batrouni_exact_2008,casula_quantum_2008,feiguin_pairing_2007,rizzi_fulde-ferrell-larkin-ovchinnikov_2008_short,tezuka_density-matrix_2008}
which are deemed to provide extra signatures of FFLO physics. A similar
analysis can be also performed for repulsive interactions between
particles, where the coexistence of different Luttinger-liquid regimes
has been proposed~\cite{kakashvili_signatures_2008}. The above approach
is well suited to explore interesting crossover effects, which are
challenging to capture with other methods. Such a system would also
be within the reach of the present generation of experiments.

\begin{acknowledgments}
We would like to thank R.~G.~Hulet's group for illuminating discussions
on experiments, and acknowledge numerous discussion with L.~O.~Baksmaty,
S.~G.~Bhongale and H.~Pu. Discussions with M.~Casula and D.~S.~Weiss
on momentum distributions are also acknowledged. We acknowledge financial
support for this project from the DARPA/ARO Grant No.~W911NF-07-1-0464.
Funding from the W.~M.~Keck and Welch Foundations is also acknowledged. 
\end{acknowledgments}
\bibliographystyle{apsrev}
\bibliography{SpinImbalancedFermions}
\end{document}